\documentclass{openjournal}
\usepackage{natbib}
\usepackage{bm}
\renewcommand{\it}[1]{\textit{#1}}
\usepackage[dvipsnames]{xcolor}
\usepackage{graphicx}	% Including figure files
\usepackage{amsmath}	% Advanced maths commands
\usepackage[linktocpage=true]{hyperref}
\usepackage{setspace}
\usepackage{verbatim}
\usepackage{makecell}
\usepackage{changepage}
\usepackage[toc,page]{appendix}
\usepackage{lineno}
\renewcommand{\arraystretch}{2}

\usepackage[xindy]{glossaries-extra}
\makeglossaries
\GlsXtrEnablePreLocationTag{Page }{Pages }

\definecolor{vlgray}{RGB}{245,245,245}
\definecolor{primary}{RGB}{255,140,0}
\definecolor{bg}{RGB}{250,250,250}
\definecolor{lgreen}{RGB}{184, 209, 126}
\definecolor{dgreen}{RGB}{112, 163, 0}
\definecolor{lorange}{RGB}{255, 199, 128}
\definecolor{dorange}{RGB}{255, 145, 0}
\definecolor{lblue}{RGB}{111, 181, 189}
\definecolor{dblue}{RGB}{0, 95, 117}

\hypersetup{
    colorlinks=true,
    linkcolor=dblue,
    citecolor=NavyBlue,
    urlcolor=lblue,
    pdftitle={Non-thermal SZ},
    pdfpagemode=FullScreen,
    }

\shorttitle{Non-thermal SZ}
\shortauthors{}
\begin{document}
\title{Sunyaev-Zeldovich signatures from non-thermal, relativistic electrons using CMB maps}

\author{\vspace{.3cm}Sandeep Kumar Acharya}

\affiliation{$^1$Astrophysics Research Center of the Open University, The Open University of Israel, Ra'anana, Israel}

\thanks{Corresponding Author: \href{mailto:sandeepkumaracharya92@gmail.com}{sandeepkumaracharya92@gmail.com}}

\begin{abstract}
Relativistically energetic, non-thermal population of electrons can give rise to unique CMB spectral distortion signatures which can be significantly different from thermal Sunyaev-Zeldovich signal or $y$-distortion. These signatures depend upon the spectrum of non-thermal electrons, therefore, a detection can inform us about the existence and abundance of non-thermal electrons in our universe. Using public CMB maps and data, we derive upper limits on non-thermal $y$-parameter for a relativistic, power-law electron distribution. With future CMB experiments, we may be in a position to detect or put significantly tighter constraints on these signals which can affect our understanding of non-thermal electron distributions in our universe.

\end{abstract}

\maketitle

%------------------------------------------------
\section{Introduction}
%------------------------------------------------
The Sunyaev-Zeldovich (SZ) effect \cite{ZS1969} has become an indispensable tool for Cosmology as well as astrophysics in recent times due to the precise observation of Cosmic Microwave Background (CMB) radiation \cite{wmap2011,PlanckSZ_2016,SPT2022,ACT2023}. The SZ effect can be of two types, (i) thermal SZ effect \cite{ZS1969},  which is produced due to the inverse Compton scattering of CMB photons by hot, energetic electrons, (ii) kinetic SZ effect \cite{SZ1980}, due to Doppler boosting by the bulk flow of electrons. The thermal SZ effect is predominantly produced at $z\lesssim 2$, once massive structures such as galaxy clusters form \cite{KS2002}. Due to the bulk motion of these galaxy clusters, they also produce the kinetic SZ effect, however, it is subdominant compared to the thermal SZ effect by an order of magnitude \cite{SBDH2010}. During reionization epoch at $6\lesssim z\lesssim 10$, formation of ionized bubbles also contribute to the total kinetic SZ effect \cite{GH1998}. The readers are referred to \cite{B1999,M2019} for a detailed review on the SZ effect. 
\par
\hspace{1cm}
Gravitational collapse of matter leads to conversion of potential energy to thermal energy which heats up the electrons to $\sim$ keV temperatures inside massive objects such as galaxy clusters.
CMB photons while passing through these structures get upscattered with probability given by the optical depth or $ n_e\sigma_{\rm T}R$ where $n_e$ is the electron density, $\sigma_{\rm T}$ is the Thomson cross section and $R$ is the line of sight dimension of the object. On each collision, the thermal electrons impart a fraction of their energy ($\sim \frac{k_{\rm B}T_{\rm e}}{m_{\rm e}{\rm c^2}}$) to the CMB photons, where $T_{\rm e}$ is the electron temperature. This effect distorts the CMB blackbody \cite{Fixsen1996} with a distinct spectral shape which is classically termed as $y$-distortion \cite{ZS1969}. This characteristic shape help us in distinguishing the thermal SZ signature from kinetic SZ as well as CMB and other astrophysical foregrounds.   
\par
\hspace{1cm}
The spectral distortion shape is sensitive to the kinematics of Compton scattering between the electron and the CMB photons.
For $T_{\rm e}\lesssim$ 1 keV, the thermal SZ signature is independent of the temperature as Compton scattering is non-relativistic. At higher temperatures, relativistic corrections become important due to Klein-Nishina terms \cite{IKN1998,Challinor1998,SS1998}.  These computations can be done extremely fast using the public code {\tt SZpack} \cite{CNSN2012}. For a non-thermal relativistic electron distribution, perturbative corrections terms are not sufficient and one needs to resort to a more general formulation \cite{EK2000,C2008}. The non-thermal SZ effect carries distinct information of the electron spectrum which produced it in the first place. Therefore, this is an ideal tool to study the abundance of relativistic electrons in our universe using CMB maps. Although, in this paper, we only consider scattering of electrons with the CMB photons, in general, these electrons will scatter all photons in the way which need not be just the CMB photons. Recent works have discussed scattering of Cosmic Infrared Background \cite{SHB2022,AC2023} and radio background \cite{HC2021} photons with the electrons residing inside galaxy clusters which can give rise to interesting signatures. 
\par
\hspace{1cm}
Possible sources of non-thermal electrons include cosmic ray electrons and relativistic electrons in radio galaxy lobes \cite{FR1969,S1974,BC1989,N1995,ZPP2014}. This electron distribution is typically a power law \cite{CMBM2013} which is inferred from radio and X-ray observations \cite{ITGHNHSM2009}. For a relativistic electron with Lorentz factor $\gamma$, a photon with energy $E_{\gamma}$ is boosted to $\gamma^2E_{\gamma}$ via inverse Compton scattering \cite{BG1970}. For electrons with $\gamma\lesssim 10$, the boosted CMB photons still show up in the CMB band ($\sim 100-500$ GHz). Electrons in the same population but with $\gamma\sim 10^4-10^5$ can boost the CMB photons to X-ray bands. These electrons in the extreme high energy tail also produce radio signature via synchrotron emission. Therefore, a detection of non-thermal SZ signature will nicely complement the radio and X-ray probes of relativistic electrons. Even a non-detection can put interesting upper limits, rule out some parameter space and can constrain our understanding of particle acceleration processes. In a recent work \cite{MB2024,CCAT2023}, the authors claim to obtain competitive constraints on magnetic field inside clusters hosting radio halos by constraining the abundance of non-thermal electrons.
\par
\hspace{1cm}
In this paper, we aim to constrain non-thermal SZ signal inside the galaxy clusters for a few representative electron spectra using {\it Planck} full sky CMB data \cite{Pl2016}. Using a stacking approach, we obtain upper limits on the amplitude of the signal. While the current limits may be uninteresting, with future CMB experiments, we may be in a position to detect or set more interesting upper limits on these signals. This paper is organized as follows. In Sec. \ref{sec:ntsz}, we describe non-thermal SZ signature in detail. We discuss the datasets used and matched filtering method in Sec. \ref{sec:data} and \ref{sec:fitering} respectively. We detail our modelling of data in Sec. \ref{sec:modelling} and explain our results in Sec. \ref{sec:results}. We conclude with some potential future directions in Sec. \ref{Sec:conclusions}. 

%------------------------------------------------------------

\section{Sunyaev-Zeldovich (SZ) effect from non-thermal electron distribution}
\label{sec:ntsz}
%------------------------------------------------------------

%----------------------------------------------------------
\begin{figure}
\centering 
\includegraphics[width=\textwidth]{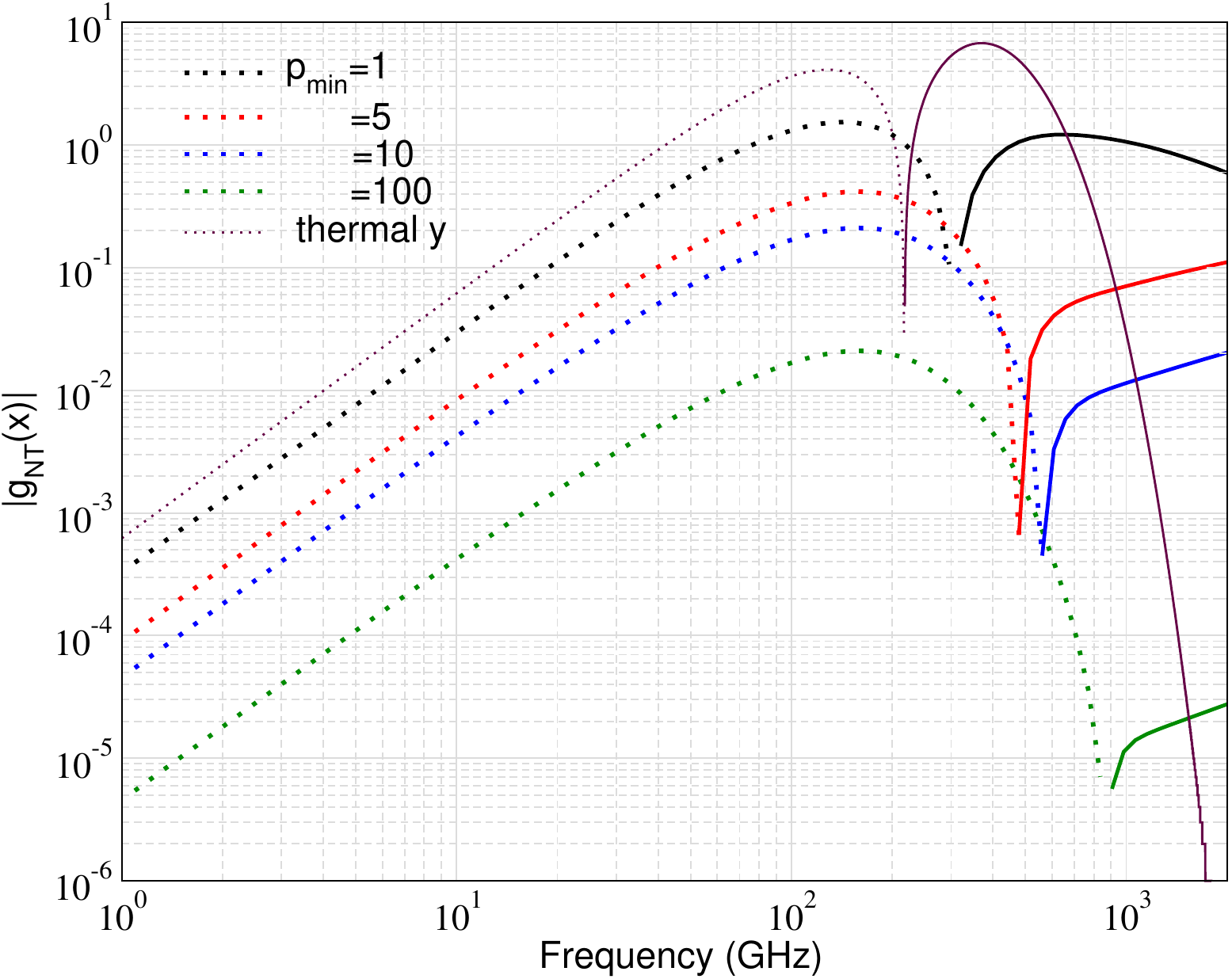}
\caption{Spectral distortion shape $g_{NT}(x)$ with $x=\frac{{\rm h}\nu}{k_{\rm B}T_{\rm cmb}}$ and for different $p_{\rm min}$. The spectral index $\alpha_p$ is assumed to be 3. The dotted and solid lines correspond to negative and positive part of $g_{NT}(x)$ respectively. We show thermal $y$-distortion for reference. }
\label{fig:spectral_shape}
\end{figure}
%------------------------------------------------------------

The CMB photons while passing through a distribution of energetic electrons can get upscattered by the electrons which imprints a characteristic distortion signal on top of CMB. For a distribution of thermal electrons, the CMB distortion signal is given by the well-known $y$-distortion \citep{ZS1969}. The distortion signal is sensitive to electron energy distribution as it depends upon the kinematics of scattering process involved which in this case is inverse Compton scattering. 
     The number of CMB photons which get upscattered from energy $x'$ to $x$ is given by,
\begin{equation}
\frac{{\rm d}N(p,x'->x)}{{\rm d}x'}= P(t,p)\times 2\frac{(k_{\rm{B}}T_{\rm{cmb}})^2}{({hc})^2}\frac{x'^2}{e^{x'}-1}, 
\label{number}
\end{equation}
where $P(t,p)$ is the kernel of the inverse Compton scattering which captures the kinematics of photon scattering with the electrons with momentum $p$, $x$ is the dimensionless frequency of photons which is given by $x=\frac{E_{\gamma}}{k_{\rm{B}}T_{\rm{cmb}}}$, $T_{\rm cmb}$ being the CMB temperature and $t=\frac{x}{x'}$. The kernel satisfies the relation, $\int {\rm d}t P(t,p)=1$, since Compton scattering is a photon number conserving process. The expression for $P(t,p)$ is given by \citep{EK2000},
%---------------------------------------------------------
\begin{equation}
P(t;p)=\frac{-3\left|(1-t)\right|}{32p^6t}[1+(10+8p^2+4p^4)t+t^2]+\frac{3(1+t)}{8p^5}\left[\frac{3+3p^2+p^4}{\sqrt{1+p^2}}-\frac{3+2p^2}{2p}(2\text{arcsinh}(p)-\left| \ln{t}\right|)\right],
\end{equation}
%---------------------------------------------------------
Exact expressions for Compton scattering kernels across all energy ranges can be found in \cite{SCL2019}.
By taking into account the number of photons which escape or enter into a particular energy bin and summing over all $x$ bins, we obtain the spectral shape of distorted CMB. This will be the case for a monochromatic electron distribution. A general distribution can be thought of as a superposition of monochromatic electron distributions with weights, for example, the differential number density of electron as a function of $p$. Then, we have to sum over the electron distribution to obtain the final spectral shape. The intensity of the distortion shape is given by,
%--------------------------------------------------------
\begin{equation}
   I_{NT}=y_{NT}g_{NT}(x),
   \label{eq:ntsz}
\end{equation}
%---------------------------------------------------------
where $I_{NT}$ is the change in intensity of distorted CMB with respect to the undistorted Planckian spectrum.
The amplitude of distortion signal is given by $y_{NT}$ (where $NT$ stands for non-thermal) whose expression is given by,
%--------------------------------------------------------------------
\begin{equation}
y_{NT}=\int \frac{p_c}{m_{\rm e}{\rm c}^2}\sigma_{\rm T}{\rm d}R,
\end{equation}
%--------------------------------------------------------------------
where $p_c$ is the red non-thermal electron black pressure and ${\rm d}R$ is a length element along the line of sight. red For equivalent thermal expression, the thermal pressure which is proportional to electron temperature has to be substituted, in the expression above black. The spectral shape $g_{NT}(x)$ is the spectral distortion shape from a normalized electron distribution and as explained above is sensitive to the non-thermal electron distribution. In this work, we assume the non-thermal electrons to have a power-law shape with electron number density $f_e(p)\propto p^{-\alpha_p}$ where $p$ is the dimensionless momentum with $p=\sqrt{\gamma^2-1}$, $\gamma$ being the Lorentz factor. We assume the spectrum to be extended from $p_{\rm min}$ to $p_{\rm max}\sim 10^6$ though reasonable change to $p_{\rm max}$ will not affect results significantly. We fix $\alpha_p$ to be 3 which is the average spectral index inferred from a population of radio galaxies \cite{CMBM2013}.   

In Fig. \ref{fig:spectral_shape}, we plot the dimensionless spectral distortion shape for a few different $p_{\rm min}$ and assuming $y_{NT}=1$. With increase in electron energy, an electron can boost the CMB photons to higher energy, therefore, moving the null point to the right. Since $y_{NT}$ captures the fraction of energy injected to the scatterd CMB photons with respect to unscattered photons, keeping $y_{NT}$ fixed and increasing $p_{\rm min}$, reduces the amplitude of $g_{NT}$. This is beacuse a higher energy electron can impart a larger fraction of energy to the CMB photons in one scattering. Therefore, as the photons move to higher energy, their overall amplitude in the CMB band reduces. It is interesting to note that below $\nu\sim 200$ GHz, the distortion shapes looks like scaled version of each other. This is expected since the target photon distribution, which is the CMB, is same for all the electrons. Therefore, the negative part of distortion or removal of photons mimics the spectrum of CMB. This also illustrates a non-trivial result that one can not distinguish different non-thermal electron distribution using multifrequency CMB experiments with frequency channels below $\sim 200$ GHz and higher frequency channels are critical for such science case. 

We should also point out that the intensity distortion of CMB ($ I_{NT}$) is the physical observable. Therefore, a detection or non-detection of non-thermal spectral distortion will result in constraints which are degenerate combination of $y_{NT}$ and $g_{NT}$. This is opposed to the case of thermal distortion for which one assumes the distortion shape to be so-called $y$-distortion \citep{ZS1969} which is universal and has no parameter dependence in it. Therefore, one can directly translate the constraints on the intensity to thermal $y$-parameter, $y_T$, or just $y$ which is the notation used in literature. However, things will not be that simple once we take into account relativistic corrections which has a dependence of gas temperature \citep{Wright1979,Fabbri1981,Rephaeli1995,Challinor1998,Sazonov1998,Itoh98} and is relevant for the environment inside the biggest galaxy clusters in the universe \citep{RBRC2019}. Then, one needs the temperature information to get constraint on thermal $y$-parameter. 

In this work, we choose a few representative electron populations with $p_{\rm min}=1$, 10 and 100. As can be noted from Fig. \ref{fig:spectral_shape}, chances of detecting non-thermal SZ signature goes down with increasing $p_{\rm min}$. However, such cases will generate signals in radio, X-ray or may be even gamma ray bands which can provide complementary constraints. Therefore, a non-detection in CMB bands while a correlated detection in higher energy bands can constrain our understanding of particle acceleration processes.

%-------------------------------------------------
\subsection{Sources of non-thermal electrons inside galaxy clusters}
%-------------------------------------------------
Compression of matter due to gravity during collapse of matter or structure formation, releases energy which heat gas. Since structure formation is a complex process, it is not impossible to believe that a significant portion of electrons do not follow thermal or Maxwell-Boltzmann distribution and remain relativistic with a power-law like spectrum. Besides, cosmic ray electrons and emission of relativistic electrons from accreting supermassive black holes can produce a non-negligible population of non-thermal electrons \cite{EK2000,C2008,CMBM2013}.  Non-thermal electrons in radio galaxy lobes produce radio emission through synchrotron emission as well as X-rays via inverse Compton process on the CMB photons \cite{CHHBBW2005,ITGHNHSM2009}. Therefore, a distortion in CMB bands is expected from the same population of electrons. Studying the correlated signatures of these non-thermal electrons across CMB, radio and X-ray bands can potentially improve our knowledge of these energetic processes significantly. In this paper, we assume that all non-thermal electrons are hosted within the galaxy clusters and use the available CMB data to constrain their abundance.

%--------------------------------------------------
\section{Datasets}
\label{sec:data}
%--------------------------------------------------

%--------------------------------------------------
\subsection{{\it Planck} full-sky maps}
%--------------------------------------------------
In this work, we have used 2nd data release of {\it Planck} full sky map \cite{Pl2016}. We have used the maps for 5 frequency channels (70,100,143,217 and 353 GHz) and ignored the lower and higher frequency channels. The lower frequency channels (30 and 44 GHz) have much lower angular resolution, 32 and 27 arcmin respectively, and are expected to be contaminated by radio point sources \citep{Planckmaps_2014,Planckmaps_2016}. Similarly, 545 and 857 GHz are severely contaminated by dust and infrared point sources. The resolution of {\it Planck} frequency maps are tabulated in Table \ref{tab:resolution}. The maps are provided in {\tt HEALPIX} \citep{healpix2005} format with $N_{\rm side}=2048$. We have reduced the maps to $N_{\rm side}=1024$ for computational simplicity. The maps between 30-353 GHz are provided in the units of ${\rm K_{\rm cmb}}$ and we have converted to the units of ${\rm MJy/sr}$ using the transformation \citep{Rybicki1979,Planck2013_spectral},
%-------------------------------------------------
\begin{equation}
    I_{\nu}=\frac{2{\rm h}\nu^3}{\rm c^2}\frac{xe^x}{(e^x-1)^2}\frac{\Delta T}{T_{\rm cmb}},
\end{equation}
%-------------------------------------------------
where $x=\frac{{\rm h}\nu}{k_BT_{\rm cmb}}$ and the maps are provided in the format of $\frac{\Delta T}{T_{\rm cmb}}$ K. We have also used the {\tt COM-Mask-CMB-IQU-common\\-field} mask, in order to reduce the Galactic contamination. {\it Planck} uses several pipelines to produce component maps and mask out the contaminants \citep{Planck2018_component}. A common mask is produced from these different pipelines, which we apply to the temperature maps, in this work.  

%-------------------------------------------------
\subsection{Galaxy cluster catalogue}
%-------------------------------------------------
Our goal of this work is to constrain non-thermal SZ contribution which we assume to be sourced inside the massive structures in the universe such as the galaxy clusters. Therefore, we use the {\it Planck} second Sunyaev-Zeldovich source Union catalog \cite{PlanckSZ_2016} which provides a sample of  1648 candidate galaxy cluster objects. The Union catalogue combines the result of three independent galaxy cluster extraction algorithms. Several of properties such as angular coordinates, redshift, mass are listed in the catalogue, however, in this work, we use only the angular coordinates of these galaxy clusters. Our use of Galactic mask reduces the number of galaxy clusters unaffected by the mask which turns out to be  1398. We use all of these objects in the mask-free region for our stacking computation.

%------------------------------------------------
\begin{table}
  \begin{center}
   \begin{tabular}{l|c|r} % <-- Alignments: 1st column left, 2nd middle and 3rd right, with vertical lines in between
      %\textbf{V} & \textbf{Value 2} & \textbf{Value 3}\\
    Frequency (GHz) & FWHM (arcmin) \\
    \hline
    70 & 13.31 \\
    100 & 9.66 \\
    143 & 7.27  \\
    217 & 5.01 \\
    353 & 4.86 
   \end{tabular}
   \caption{Characteristics of {\it Planck} maps used in this work.}
   \label{tab:resolution}
   \end{center}
%   \label{tab:resolution}
   \end{table}
   %-----------------------------------------------

%-------------------------------------------------
\section{Matched filtering method}
\label{sec:fitering}
%-------------------------------------------------
In this work, we have used the matched filtering approach to extract the temperature fluctuation along the direction of the galaxy clusters. The matched filtering approach was developed in the works of \cite{HT1996,HSHBDML2002,MBD2006} and was subsequently applied to the CMB data to extract the galaxy cluster catalogue, as an example see \cite{PlanckSZ_2016}. For more recent and interesting applications, the readers are referred to \cite{SGEPS2017,EBCB2018}. 
In this approach, we use a template for the signal we are looking for, in order to, extract this signal from a map containing the signal along with foregrounds and noise. In general, one can use both a spatial and frequency template to make the filtering process more effective. Indeed, it is typically assumed that the hot gas inside the galaxy clusters give rise to thermal SZ signal predominantly and its spectral independence from all other foregrounds along with CMB as well as the spatial profile of clusters is used to identify the galaxy cluster candidates. Here, we use only the spatial information in the filtering step for individual galaxy cluster candidate and use a frequency template to isolate non-thermal SZ signal only after stacking all the objects in our galaxy catalogue.   

The signal at an angular location $x$ in sky can be modelled as,
%-------------------------------------------------
\begin{equation}
    M(x)=s_0T_{\theta_c}(x-x_0)+N(x),
\end{equation}
%-------------------------------------------------
where $T_{\theta_c}$ is the spatial template for intensity at a single frequency, parameterized by $\theta_c$, where $\theta_c$ is the radius of our filter profile and centered at $x_0$, $s_0$ is the magnitude of the signal component, that we are after, at the center of the object and $N(x)$ is the noise which refer to every other signal that we are not interested in including instrumental noise. We want to construct a filter which when applied to this data will reliably give us back the signal we are after or concretely,
%------------------------------------------------
\begin{equation}
    s_0=\int d^2x \Psi_{\theta_c}(x-x_0)M(x)
    \label{eq:signal_amplitude}
\end{equation}
%------------------------------------------------
For the filtering procedure, we need to transform the filter from position or pixel space to fourier space. Since, the sky is curved, the procedure involves introducing spherical harmonics. However, the effect of sky curvature is important at degree scales. Since the objects under study have a size of order few arc minutes
(See Eq. \ref{eq:theta_500} and discussions below), we can ignore this effect and can assume that the sky is flat on the relevant length scales.
In the flat sky approximation, the filter in the fourier space is given by \citep{MBD2006},
%------------------------------------------------
\begin{equation}
    \Psi_{\theta_c}(k)=\sigma_{\theta_c}^2P^{-1}(k)T_{\theta_c}(k),
\end{equation}
%-----------------------------------------------
where $\sigma_{\theta_c}=\left[\int d^2k \frac{T_{\theta_c}(k)^2}{P(k)}\right]^{-1/2}$. In the flat-sky approximation, there is a one-to-one correspondence between the angular multipole $l$ and fourier mode $k$ as $l=2\pi k$. Also the noise power spectrum can be identified with $C_l$ which can be be directly extracted from the map. Technically, the noise power spectrum extracted from the data includes the contribution from the signal that we are looking after. However, in the case of galaxy clusters, this contribution is negligible compared to the CMB as well as the foregrounds. A recent work looks into a more accurate computation using an iterative approach \cite{ZRCB2023} which can be important for future CMB experiments.  The filter is convolved with the beam since one can not resolve sources with size less than the width of the beam. Including the effect of beamwidth, the expression for filter becomes,
%----------------------------------------------
\begin{equation}
    \Psi_{\theta_c}(l)=\frac{C_l^{-1}T_{\theta_c}(l)B_l}{\left[\frac{1}{2\pi}\int ldl \frac{T_{\theta_c}(l)^2}{C_l}\right]}
\end{equation}
%-----------------------------------------------
The beam is assumed to be gaussian with FWHM given in Table \ref{tab:resolution}. This assumption simplifies the analysis and for a more realistic description of the {\it Planck} beams, the readers are referred to \cite{Planck2016_beams}.

%-------------------------------------------------
\begin{figure}
\centering 
\includegraphics[width=\columnwidth]{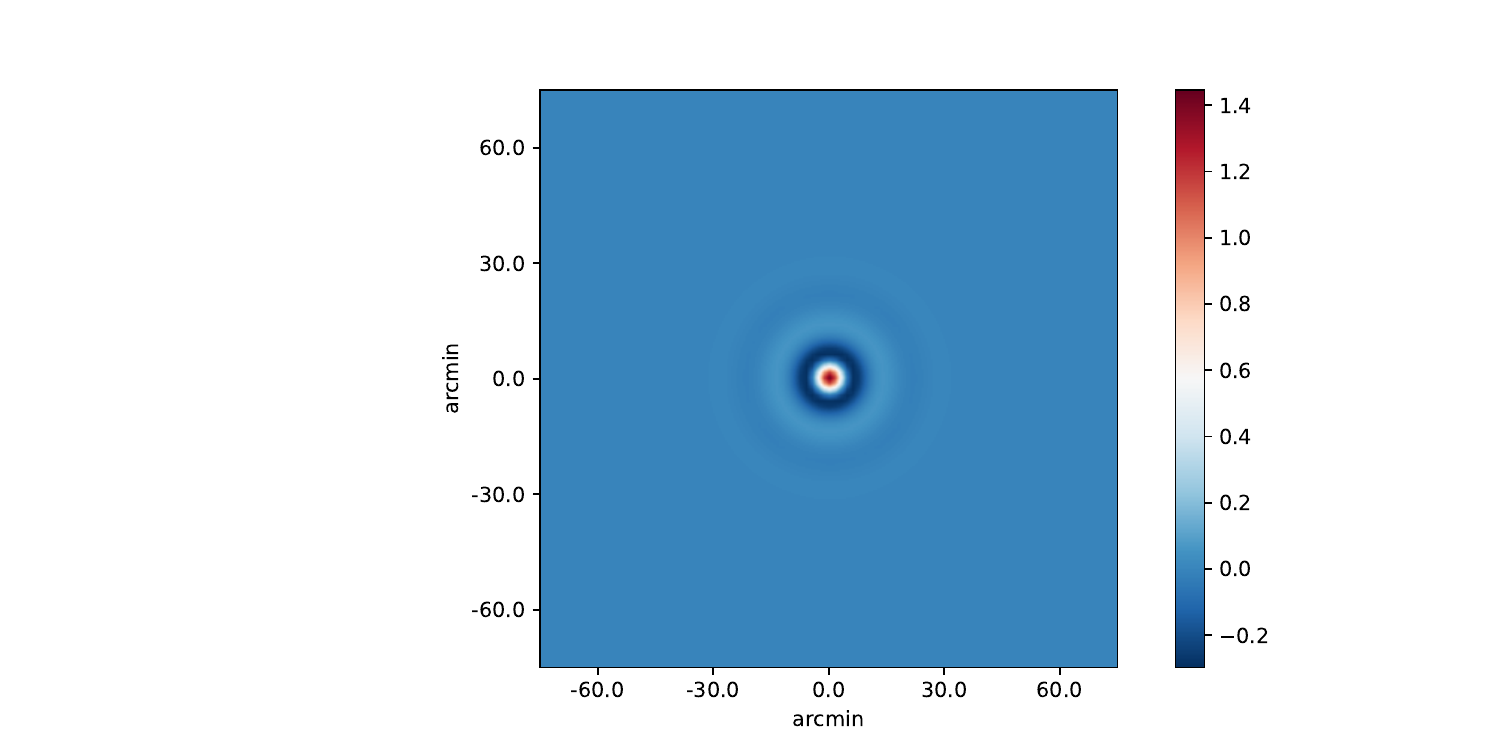}
\caption{Matched filter in position or pixel space at $\nu=143$ GHz.  }
\label{fig:filter}
\end{figure}
%-------------------------------------------------

We assume that the non-thermal SZ signature is correlated with the profile of gas in galaxy clusters which is typically assumed to be a spherical beta model (See \cite{SGEPS2017,EBCB2018}). However, for more accurate analysis, realistic profiles such as Arnaud profile \cite{Arnaud2010} or Battaglia profile \cite{Battaglia2012} should be used in future. A signal correlated with the profile of gas inside the cluster can be expressed as,
%----------------------------------------------
\begin{equation}
    s(\theta)=s_0\left[1+\left(\frac{\theta}{\theta_c}\right)^2\right]^{(1-3\beta)/2},
\end{equation}
%-----------------------------------------------
where $\theta_c$ is the core radius and $\beta$ is assumed to be 1. The analytic spherical transform of the profile with $s_0=1$ is given by \cite{SGEPS2017},
%-----------------------------------------------
\begin{equation}
   T_{\theta_c}(l) =2\pi\theta_c^2K_0(l\theta_c), 
\end{equation}
%-----------------------------------------------
where $K_0$ is the modified Bessel function of the second kind.

The galaxy clusters resides inside dark matter halos whose size are denoted by $\theta_{500}$ which can be related to its mass and redshift as,
%-----------------------------------------------
\begin{equation}
    \theta_{500}=6.997\left[\frac{h}{0.7}\right]^{-2/3}\left[\frac{(1-b)M_{500}}{3\times 10^{14}M_{\odot}}\right]^{1/3}E(z)^{-2/3}\left[\frac{D_A(z)}{\rm 500 Mpc}\right]\hspace{0.5cm} (\rm arcmin),
    \label{eq:theta_500}
\end{equation}
%-----------------------------------------------
where $M_{500}$ is the mass of the halo inside the radius of $\theta_{500}$, $E(z)=H(z)/H_0$ and $D_A(z)$ is the angular diameter distance and $b$ is the hydrostatic bias \cite{Planck2014_powerspectrum}. For the redshift and mass distribution of galaxy cluster catalogue, $\theta_{500}$ varies between $\sim 1$ to 10 arcmin. The core size is related to halo size by the expression, $\theta_c=c_{500}\theta_{500}$, where $c_{500}\sim 0.2$ using the concentration-mass relation \cite{HK2003,DSKD2008}. In this work, we fix $\theta_c=1$ arcmin to be the representative size of the galaxy clusters and do not bin the objects by their sizes. We do not expect this choice to significantly affect our results as we do a stacking computation which should average out these effects.      

In Fig. \ref{fig:filter}, we show the matched filter in pixel space which we use for the 143 GHz map. The filter is polar symmetric which is expected due to our choice of cluster profile. The filter shows ringing pattern with positive and negative values which helps in removing large scale foregrounds, uncorrelated to the galaxy clusters. Essentially, our matched filter acts as a high pass filter.
We apply the filter to appropriate frequency maps and then stack all the unmasked galaxy clusters. In principle, one can bin the clusters in mass or redshift. However, we expect the signal to be weak and binning will make it even more weaker. Therefore, in this work, we have stacked all the clusters without any binning. This is useful in providing upper limits and once we have a detection, we can look for variation of signal as a function of mass or redshift of the galaxy clusters. 

\begin{figure}
%\begin{subfigure}[b]{0.4\textwidth}
\includegraphics[scale=0.53]{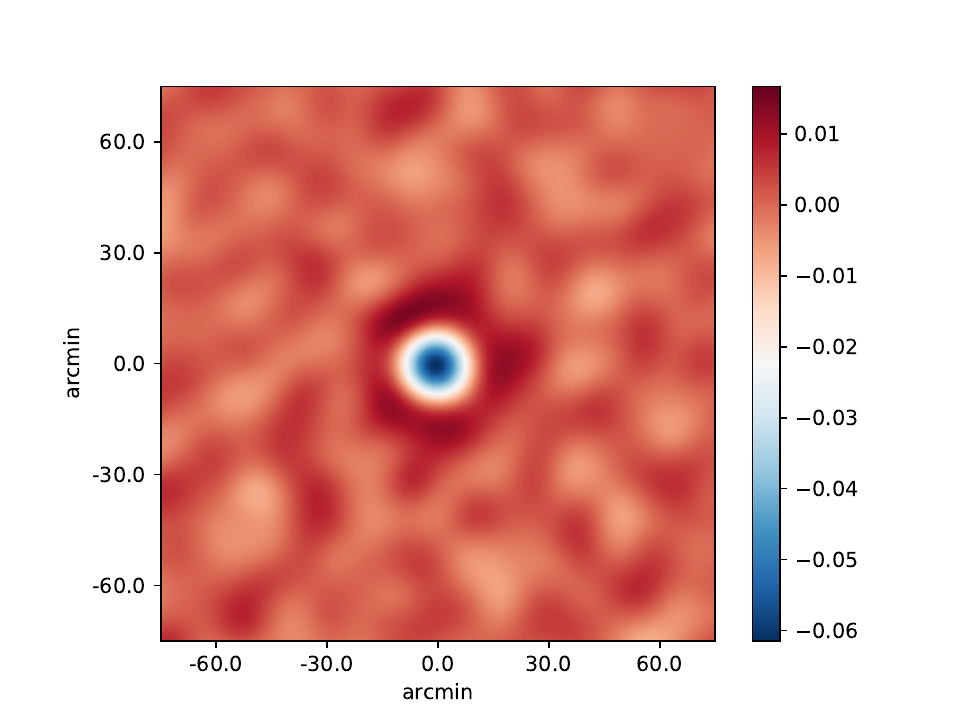}
%\caption{70 GHz}
%\label{fig:depfracz=1000}
%\end{subfigure}\hspace{50 pt}
%\begin{subfigure}[b]{0.4\textwidth}
\includegraphics[scale=0.53]{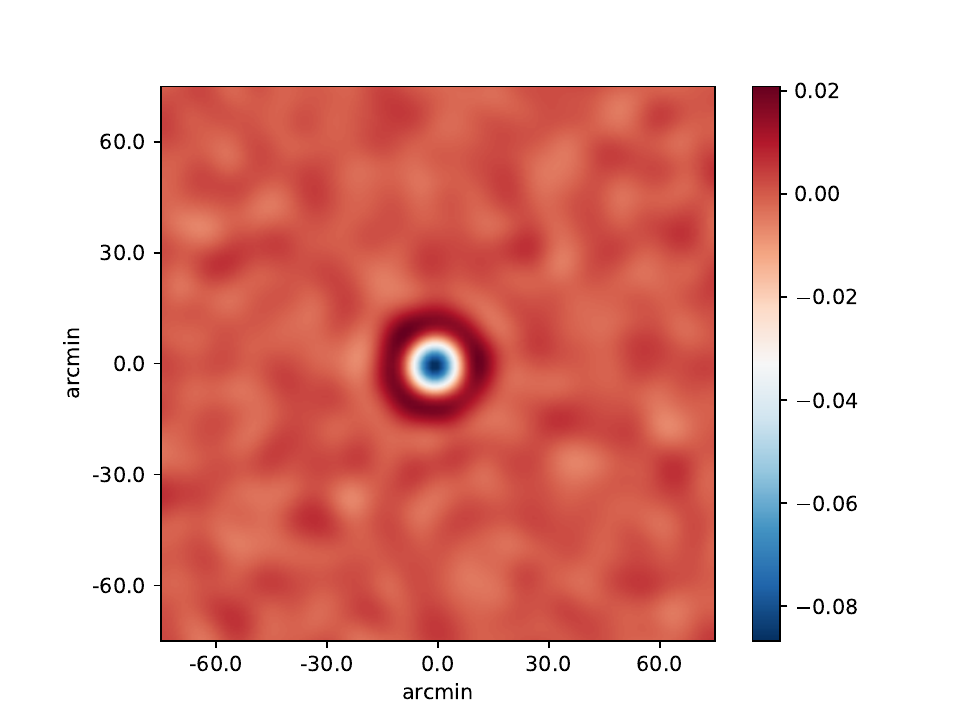}
%\caption{100 GHz}
%\label{fig:depfracz=100}
%\end{subfigure}
%\begin{subfigure}[b]{0.4\textwidth}
\includegraphics[scale=0.53]{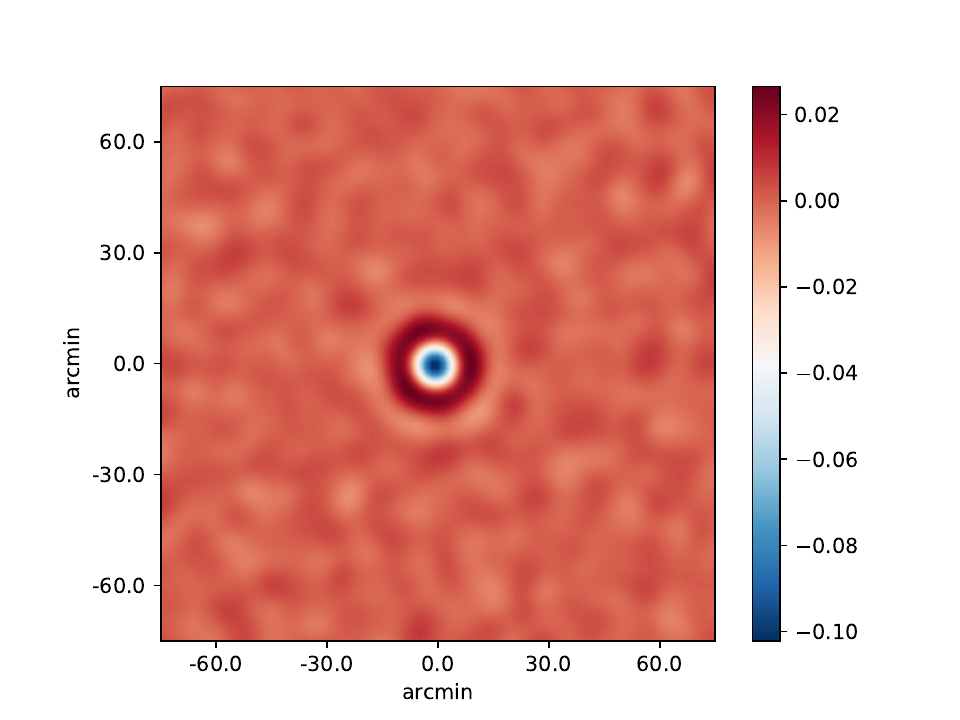}
%\caption{143 GHz}
%\label{fig:depfracz=1000}
%\end{subfigure}\hspace{50 pt}
%\begin{subfigure}[b]{0.4\textwidth}
\includegraphics[scale=0.53]{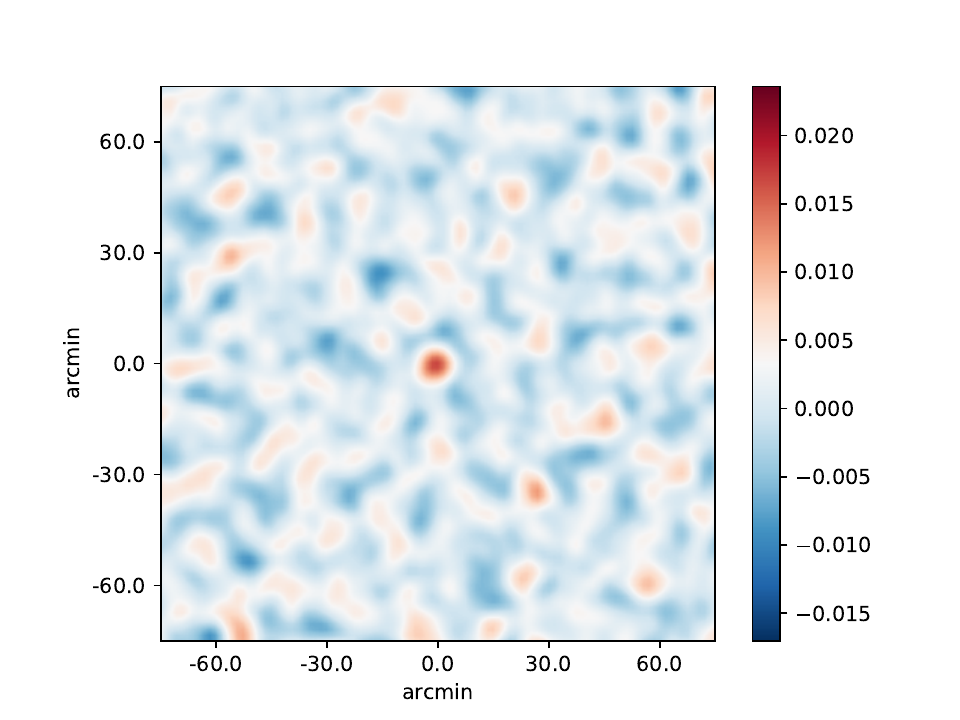}
%\caption{217 GHz}
%\label{fig:depfracz=100}
%\end{subfigure}
%\end{figure}
%\begin{subfigure}[b]{0.4\textwidth}
\includegraphics[scale=0.53]{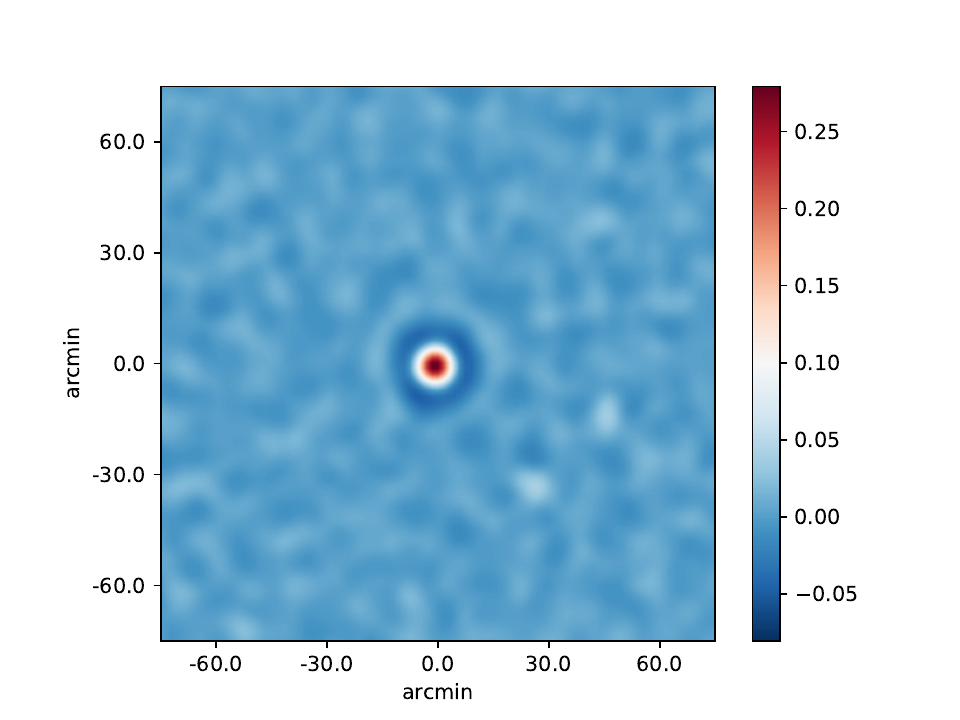}
%\caption{353 GHz}
%\label{fig:depfracz=1000}
%\end{subfigure}\hspace{50 pt}
\caption{Matched filtered, stacked intensity maps of galaxy clusters for the different frequency channels, 70 GHz (top left), 100 GHz (top right), 143 GHz (middle left), 217 GHz (middle right) and 353 GHz (bottom left). The numbers are shown in ${\rm MJy/sr}$ unit.}
\label{fig:MF_figure}
\end{figure}

In Fig. \ref{fig:MF_figure}, we show the spectral intensity of the signal at the location of clusters by stacking them. We generate a 2D map since our matched filter looks for the associated signal by scanning across the input maps which are two-dimensional. Essentially, the center of the filter moves from one pixel to another and this procedure gives a value of $s_0$ (Eq. \ref{eq:signal_amplitude}) for each pixel which leads to the filtered 2D map.  One can clearly see the stacked signal with high significance at the central pixel. One can also see that the spectral intensity is negative at $\nu<217$ GHz while positive at higher frequency. We now move to fitting frequency template to the stacked, matched filtered data and discuss our modelling in the next section.

%---------------------------------------------------

%------------------------------------------------

%-------------------------------------------------
\section{Data modelling}
\label{sec:modelling}
\begin{figure}
\centering 
\includegraphics[width=\columnwidth]{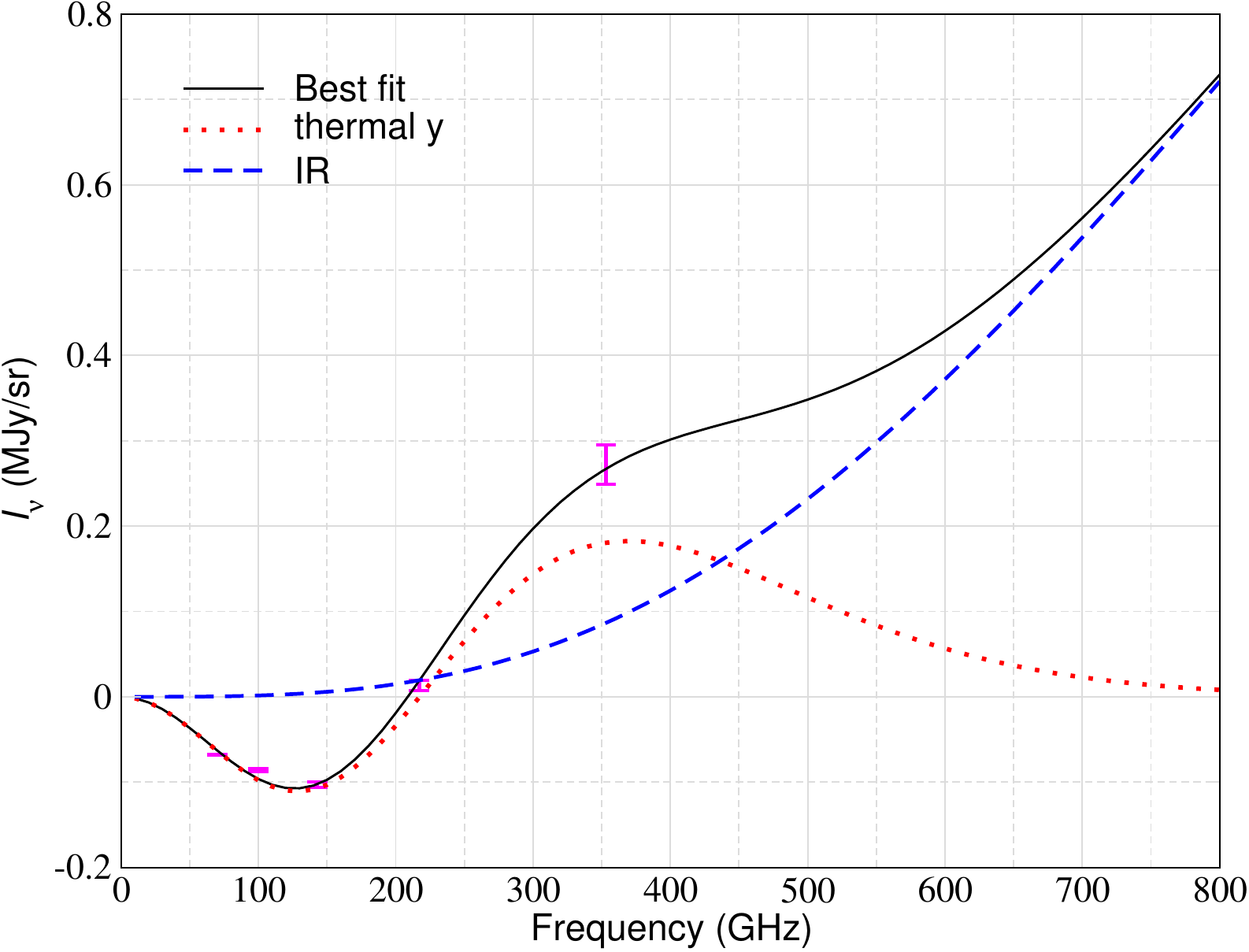}
\caption{Fit to matched filtered, stacked spectral intensity data. As a starting point, we fit thermal $y$-ditortion, IR spectrum to these data points and show the best fit combination.  }
\label{fig:MF_fit_70_353}
\end{figure}

\begin{figure}
\centering
%\begin{subfigure}[b]{0.4\textwidth}
\includegraphics[scale=0.3]{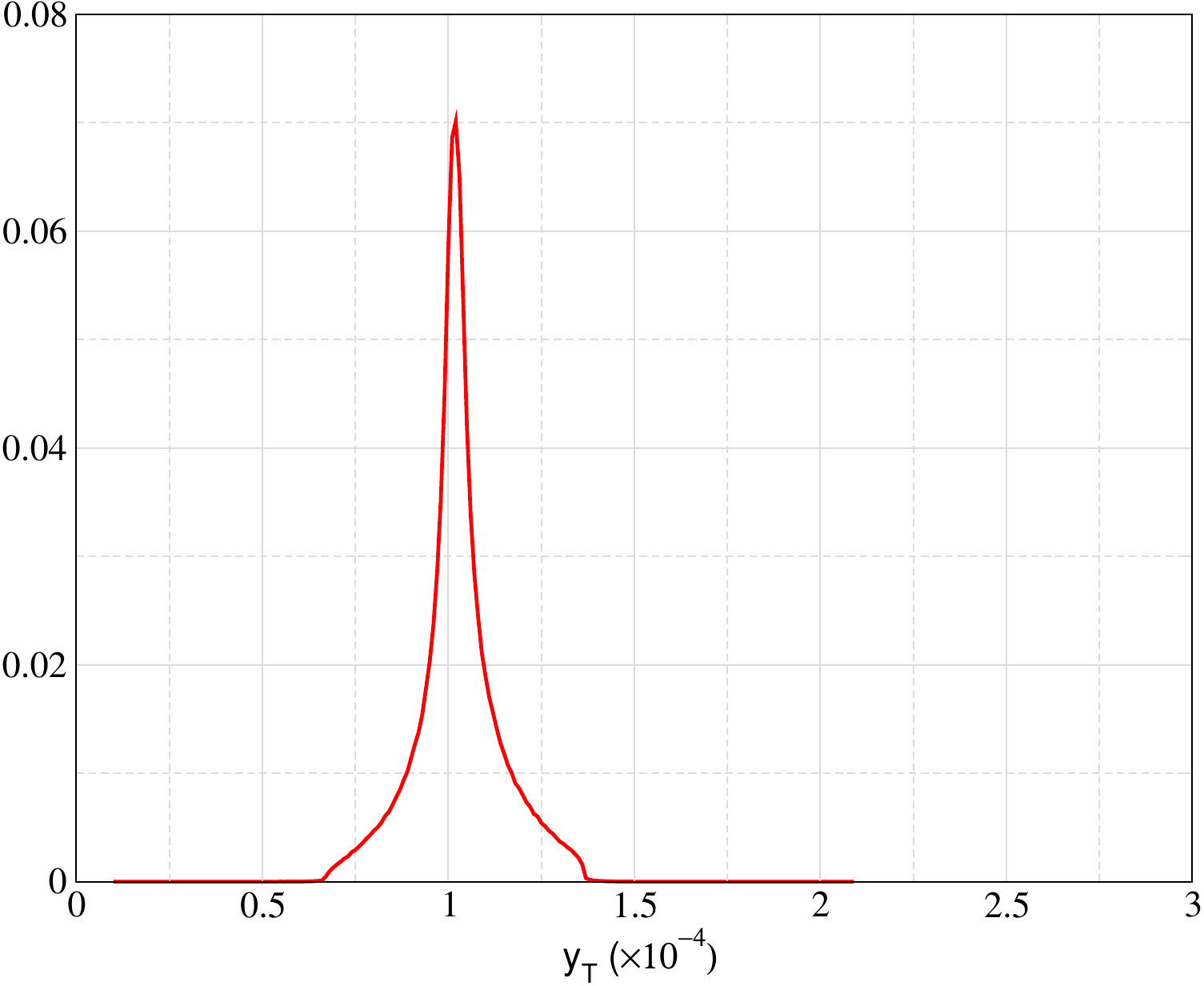}
%\caption{70 GHz}
%\label{fig:depfracz=1000}
%\end{subfigure}\hspace{50 pt}
%\begin{subfigure}[b]{0.4\textwidth}
\includegraphics[scale=0.3]{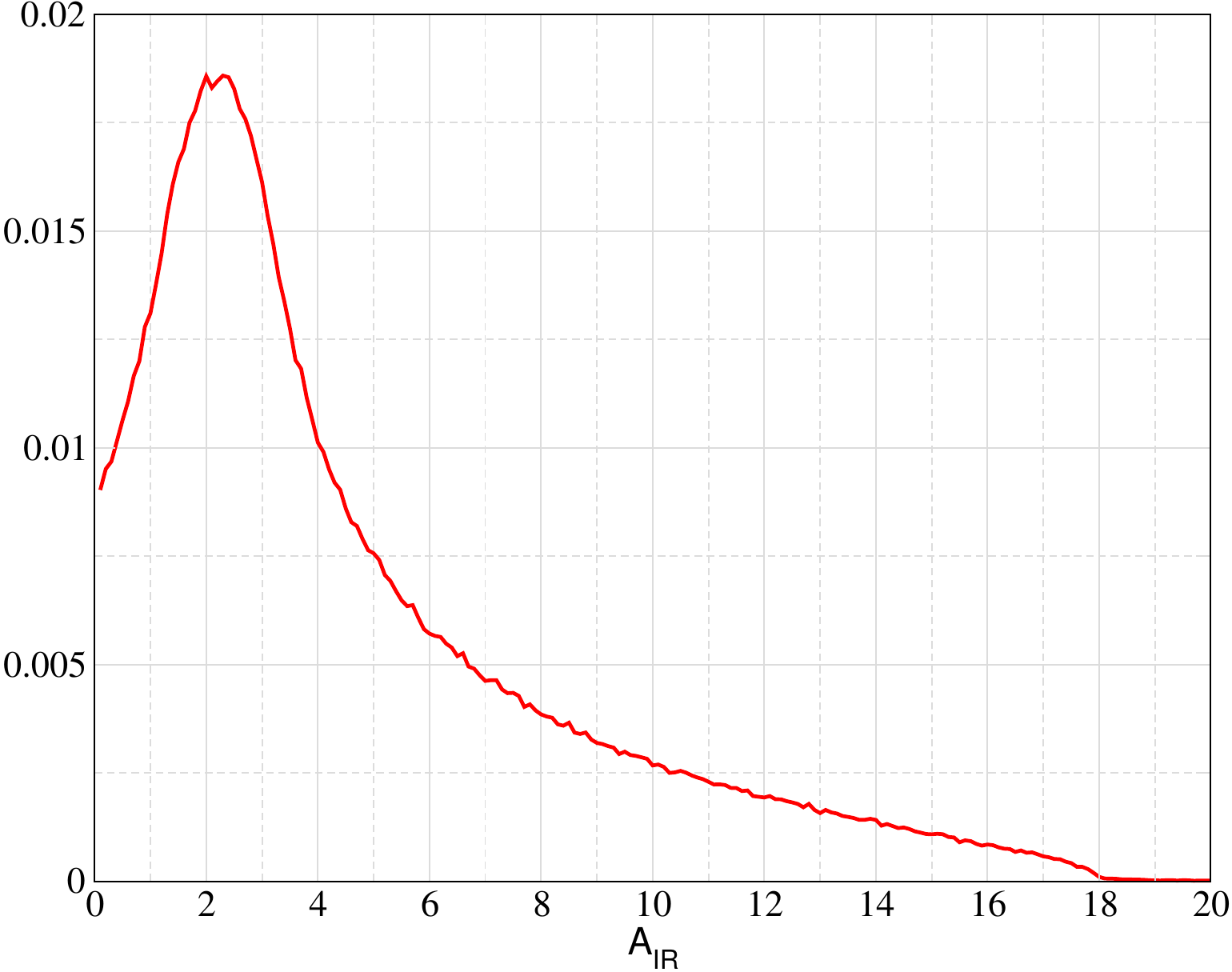}
%\caption{100 GHz}
%\label{fig:depfracz=100}
%\end{subfigure}
\caption{Marginalized constraints on $y_T$ and $A_{\rm IR}$ by fitting only thermal $y$ and IR component to the data.}
\label{fig:y_IR_const}
\end{figure}

%-------------------------------------------------
After filtering, the only surviving contribution from galaxy clusters are the thermal SZ and infrared emission correlated to the clusters, in addition to non-thermal SZ, if any. Therefore, we fit a three component model to our data. The thermal $y$-distortion is given by the analytic formula \cite{ZS1969},
%-------------------------------------------------
\begin{equation}
    I_{T}=y_T\frac{2{\rm h}\nu^3}{\rm c^2}\frac{x{\rm e}^x}{({\rm e}^x-1)^2}\left[x\frac{{\rm e}^x+1)}{{\rm e}^x-1}-4.0\right],
\end{equation}
%-------------------------------------------------
where $x=\frac{{\rm h}\nu}{k_BT_{\rm cmb}}$ and $y_T$ is the amplitude which is a free parameter. Similarly, infrared component can be accurarely described across {\it Planck} frequencies as a modified blackbody which can be written as, \citep{Planckmaps_2014, Planckmaps_2016},
%-------------------------------------------------
\begin{equation}
    I_{{\rm IR}}=A_{IR}\times 10^{-6}\frac{2{\rm h}\nu^3}{\rm c^2}\left(\frac{\nu}{\nu_0}\right)^{\beta}\frac{1}{({\rm e}^{{\rm h}\nu/k_BT_d}-1)},
\end{equation}
%-------------------------------------------------
where $A_{\rm IR}$ is a free parameter which determines the overall amplitude. 
We have fixed the value of $\nu_0=353$ GHz, $\beta=1.5$ and $T_d=18$ K which are their best fitting values \citep{Planckmaps_2014, Planckmaps_2016}. This simple choice of template and parameters fit the data well \cite{Planck_infrared} but we do not claim that it captures all the physics correctly. In \cite{SHKO2012} and \cite{MM2021}, the authors describe a halo model approach which motivate a similar spectral template within the context of current data. The non-thermal distortion is
%\clearpage
parameterized as in eq. \ref{eq:ntsz},
%------------------------------------------------
\begin{equation}
      I_{NT}=y_{NT}g_{NT}(x)
\end{equation}
%------------------------------------------------
%\clearpage
In our modelling, we allow only the amplitude of the three components to vary. To estimate the noise covariance matrix, we stack uniformly distributed random positions, with same number as many as galaxy clusters in our sample, in the map.
%\clearpage
We repeat the exercise for $10^4$ times to
obtain a distribution which sets the variance for individual frequency channel.  We note that this procedure gives a lower bound on the noise and ignores intrinsic scatter of the measurements \citep{MBCDDRB2018}.   
%\clearpage
We ignore correlation across different frequency channels, therefore, our covariance matrix is diagonal. We note that the noise which also includes CMB, correlates across the channels significantly. More details about this aspects can be found in \cite{SGEPS2017,EBCB2018}. 
\clearpage
However, we do not expect that this effect will change our constraints on non-thermal SZ by orders of magnitude. As a basic check, we obtain the magnitude of thermal SZ ($y_T$) which is shown below and is in reasonable agreement with previous calculations.      
%\clearpage
Assuming gaussian likelihood, we use Markov Chain Monte Carlo (MCMC) sampling to obtain posterier distribution on the free parameters. In Fig. \ref{fig:MF_fit_70_353}, we show our matched filtered, stacked data and best fit combination to data with just thermal $y$ and IR. In addition, we show the marginalized posterior distribution on $y_T$ and $A_{IR}$ in Fig. \ref{fig:y_IR_const}. The best fit $y_T$ and $A_{IR}$ turns out to be $\sim 10^{-4}$ and 2.2 respectively. In the next section, we include non-thermal SZ and derive marginalized constraints on $y_{NT}$.

%\clearpage
%----------------------------------------------------------
%-----------------------------------------------------------

%---------------------------------------------------
%--------------------------------------------------

%-------------------------------------------------
\section{Results}
\label{sec:results}
%-------------------------------------------------
In this section, we provide our main results in the form of upper limits on $y_{NT}$ for a few different non-thermal electron spectrum parameterized by $p_{\rm min}$ as in Sec. \ref{sec:ntsz}. We use flat prior on $y_T$, $y_{NT}$ and $A_{\rm IR}$ for MCMC sampling.  

%-----------------------------------------------------
\subsection{Constraining $y_{NT}$ with varying $y_T$ and $A_{IR}$ }
\begin{figure}
\centering
%\begin{subfigure}[b]{0.4\textwidth}
\includegraphics[scale=0.3]{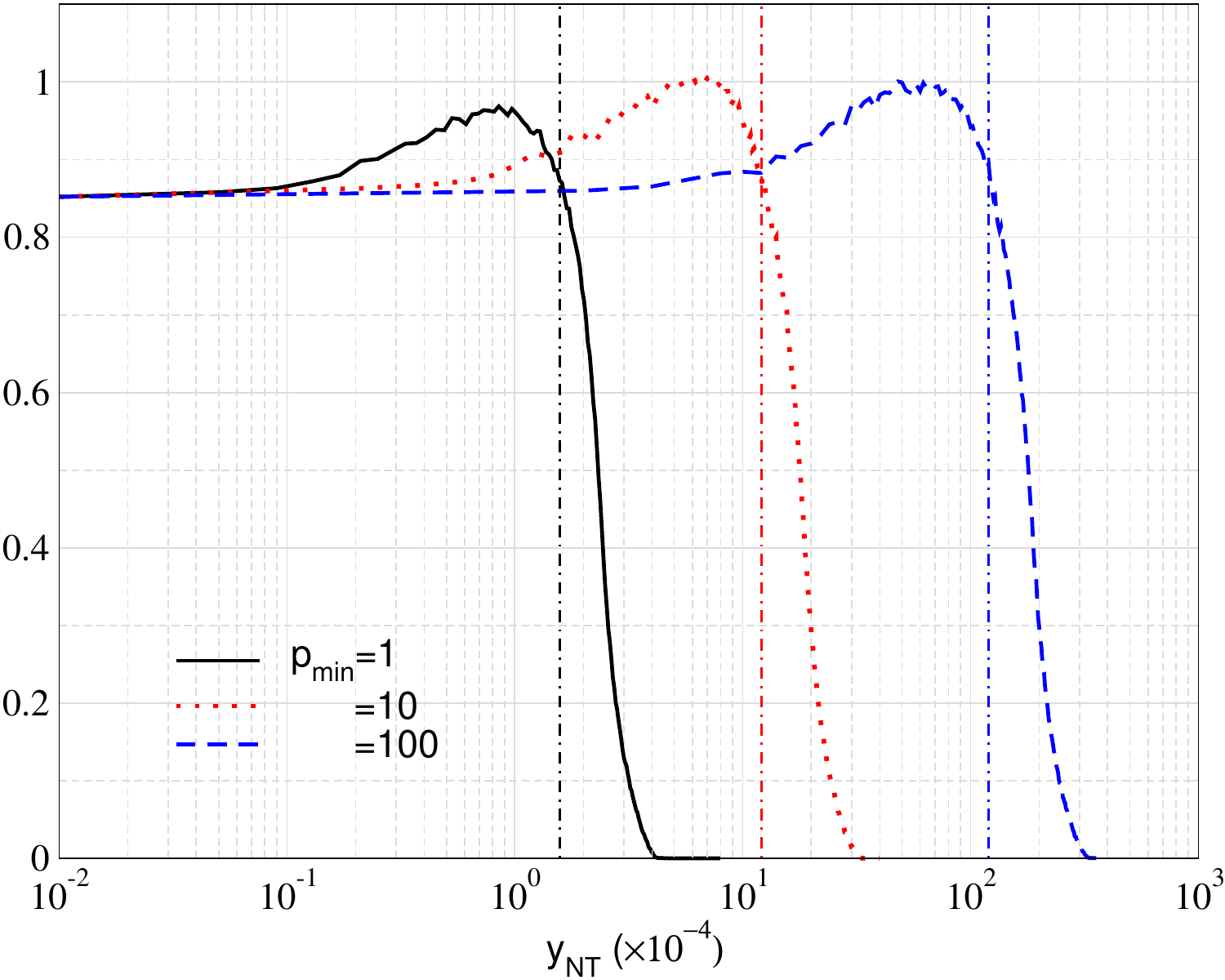}
%\caption{Marginalized posterior on $y_{NT}$ with varying thermal y, non-thermal y and IR}
%\label{fig:depfracz=1000}
%\end{subfigure}\hspace{50 pt}
%\begin{subfigure}[b]{0.4\textwidth}
\includegraphics[scale=0.3]{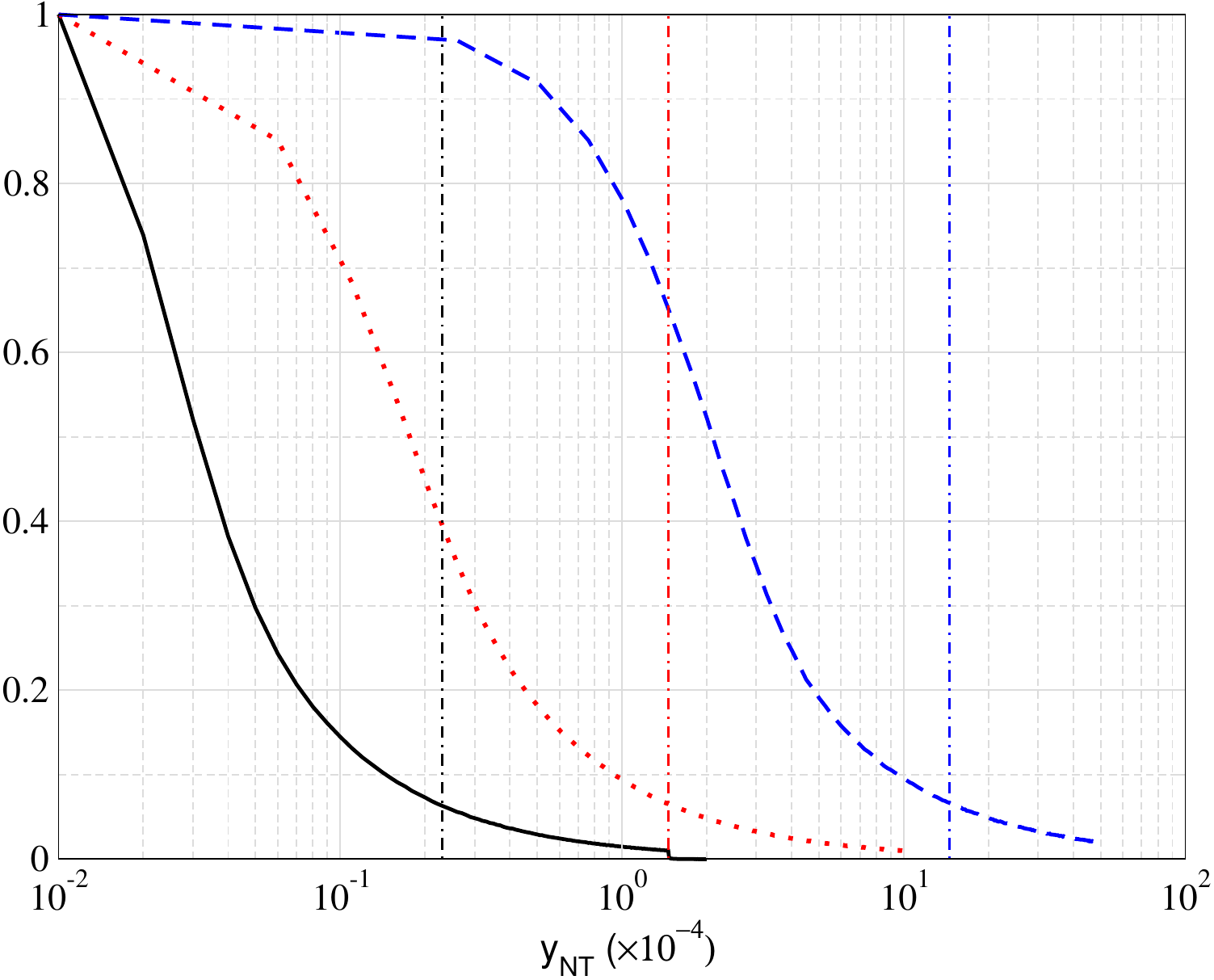}
%\caption{Marginalized posterior with fixed thermal y and IR with their best fit value.}
%\label{fig:depfracz=100}
%\end{subfigure}
\caption{(a) Marginalized posterior on $y_{NT}$ with varying $y_T$ and $A_{IR}$ (left), (b) by fixing $y_T=10^{-4}$ and $A_{IR}=2.2$ obtained in previous section (right). We show 68 percent confidence interval in dot-dashed vertical lines.}
\label{fig:y_IR_yNT_free_const}
\end{figure}

%-----------------------------------------------------
In Fig. \ref{fig:y_IR_yNT_free_const}, we show the marginalized posterior on $y_{NT}$ by simultaneously varying $y_T$, $y_{NT}$ and $A_{IR}$. For $p_{\rm min}=1$, we obtain upper limit on $y_{NT}$ which is of the order of $10^{-4}$. This is expected as the spectral shape for $p_{\rm min}=1$ is similar to thermal $y$-distortion (at least for $\nu\lesssim 200$ GHz), therefore, our upper limit is of the same order as the best fit $y_T$ as obtained in the previous section. The constraints weaken significantly for higher $p_{\rm min}$ which again follows from behaviour as seen in Fig. \ref{fig:spectral_shape}. 

We further consider a case, where we fix $y_T=10^{-4}$ and $A_{IR}=2.2$. This corresponds to the case where we have a strong prior on both of these parameters. This might be possible with future CMB experiments such as \cite{SO2019,CMB_HD2019,CCAT2023} where we will detect thermal SZ signature with very high significance and can model IR component much better with more high frequency channels. As expected, we obtain much stronger constraints with $y_{NT}\lesssim 10^{-5}$ for $p_{\rm min}=1$.

%---------------------------------------------------
%--------------------------------------------------

%-----------------------------------------------------
\subsection{Constraining $y_{NT}$ with varying $y_T$ only}
\begin{figure}
\centering
%\begin{subfigure}[b]{0.4\textwidth}
\includegraphics[scale=0.3]{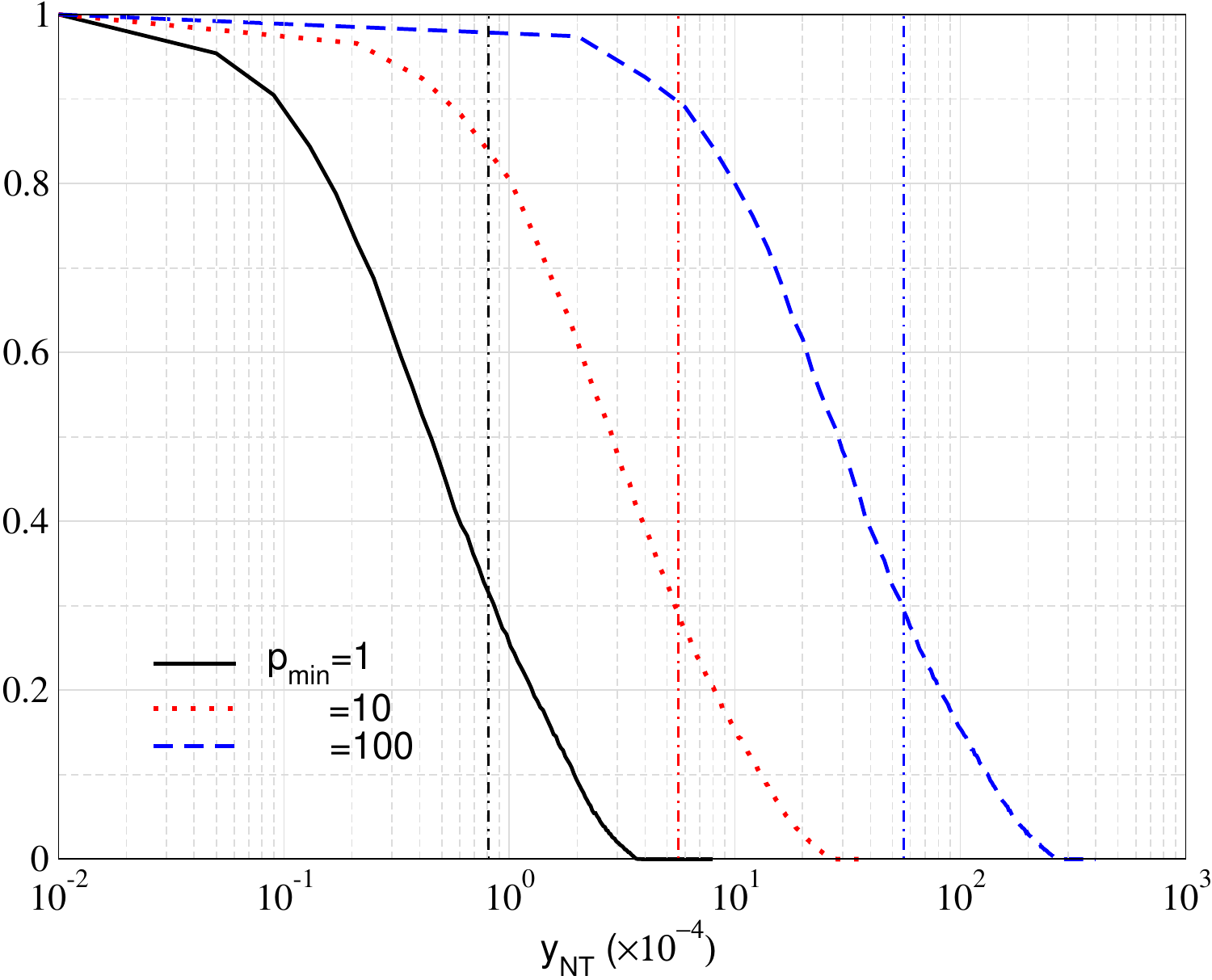}
%\caption{Non-thermal SZ constraint with varying thermal y, non-thermal y}
%\label{fig:depfracz=1000}
%\end{subfigure}\hspace{50 pt}
%\begin{subfigure}[b]{0.4\textwidth}
\includegraphics[scale=0.3]{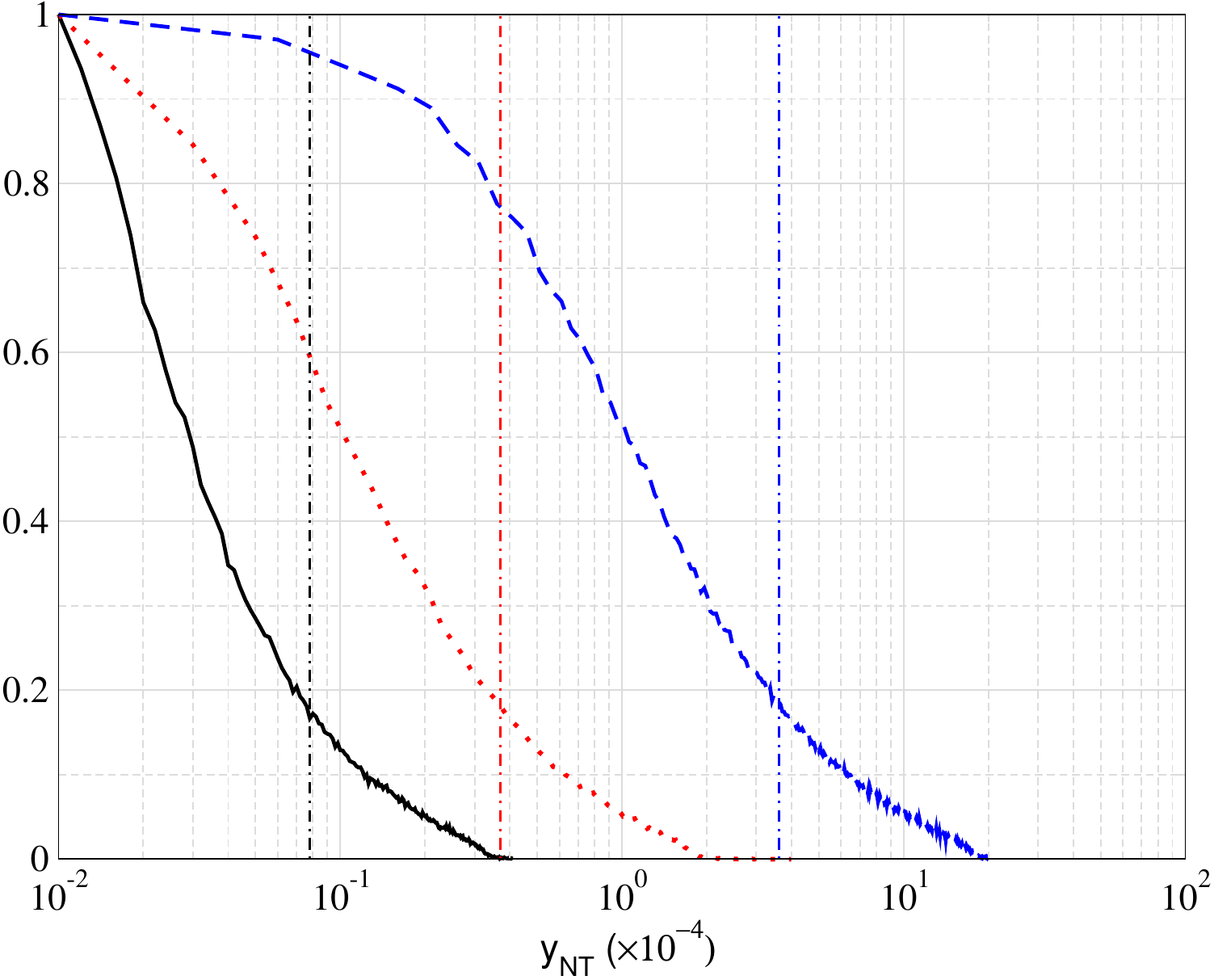}
%\caption{Non-thermal SZ constraint with varying thermal y but noise reduced by a factor of 10. }
%\label{fig:depfracz=100}
%\end{subfigure}
\caption{ (a) Marginalized posterior on $y_{NT}$ with varying $y_T$ (left), (b) same with rescaling the noise down by a factor of 10 (right).  }
\label{fig:y_only_const}
\end{figure}

%-----------------------------------------------------
As described in the previous section, we have used a simple description for IR component where we have fixed some of the parameters (dust temperature as an example). Depending on the complexity of the modelling, constraints on $y_{NT}$ obtained above might change. Therefore, in this section, we do not fit for IR component to the data. In Fig. \ref{fig:y_only_const}, we show the marginalized posterior on $y_{NT}$ for this case. We obtain upper limits which are quantitatively similar compared to the previous subsection.

In this work, we have used a sample of $\sim 10^3$ galaxy clusters while, in future, we might have a sample of $10^5$ clusters \cite{CMB_HD2019}. Assuming the noise in Fig. \ref{fig:MF_fit_70_353} goes down by $\frac{1}{\sqrt{N}}$, where $N$ is the number of galaxy clusters, the constraints can further tighten up or may lead to a potential detection.
We note that, this gives a very rough estimate of the noise, as it is not just the number of clusters that is expected to increase, but the observations for each clusters are also expected to have higher signal-to-noise. We simulate such a situation where we have simply scaled down the noise in Fig. \ref{fig:MF_fit_70_353} by a factor of 10 and repeating the analysis. The results are shown in the right hand panel of Fig. \ref{fig:y_only_const}. We find that keeping up with the noise level, the constraints also tighten up by an order of magnitude and reach a level of $y_{NT}\lesssim 10^{-5}$ for $p_{\rm min}=1$. An upper limit of $y_{NT}\lesssim 10^{-5}$ would constraint non-thermal electron pressure to less than about 10 percent of thermal electrons. Even though higher $p_{\rm min}$ results in weaker constraints, a non-detection will result in ruling out some parameter space in conjunction with radio or X-ray observations. This will have interesting implications for particle acceleration mechanisms in cosmic objects like galaxy clusters.  

%---------------------------------------------------
%--------------------------------------------------

%---------------------------------------------------
\section{Conclusion}
\label{Sec:conclusions}
%---------------------------------------------------
In this paper, we have obtained constraints on the non-thermal SZ signatures with a power-law electron population, assuming that these non-thermal electrons are hosted inside the galaxy clusters. We have used {\it Planck} public data and galaxy cluster catalogue to obtain the upper limits on non-thermal SZ for a few representative electron populations. Unsurprisingly, these upper limits depend upon the electron spectrum and our strongest upper limit is of the order of $y_{NT}\sim 10^{-4}$. This would imply that the maximum possible non-thermal electron pressure is about the same order of magnitude compared to thermal electrons. This may be physically non-interesting as thermal electrons are expected to dominate non-thermal electrons inside the galaxy clusters \citep{ZPP2014}. However, with increase in number of galaxy cluster detections with future CMB experiments such as Simons Observatory \citep{SO2019}, CMB-S4 \citep{CMBS42016}, CMB-HD \citep{CMB_HD2019}, we could tighten the upper limits by an order of magnitude. Even with eROSITA galaxy cluster catalogue \cite{erosita2012} with about $10^5$ detections, we may be able to reach that goal in near future. We also point out that for improved constraints, we need to have better control over the foregrounds as well. In this work, we have fixed the infrared spectrum for the analysis but, in reality, we have to marginalize over the foregrounds which might degrade the constraints. Most of the constraining power, in this work, comes from  $\nu\lesssim 217$ GHz channels, where infrared foreground is subdominant. Therefore, we do not expect significant changes from the constraints that we have obtained with a fixed infrared spectrum. However, it will be an important issue for upcoming CMB experiments. A different way to parameterize the foregrounds using moment expansion method \citep{CHA2017} might also be pursued as an alternate way.  While a detection will be exciting, a non-detecton of non-thermal SZ signature in combination with radio, X-ray surveys can put strong constraints on parameter space of potential non-thermal electron spectra and expand our knowledge of particle acceleration mechanism inside galaxy clusters.

%-------------------------------------------------
\section{Acknowledgements}
%-------------------------------------------------
This work is supported by ARCO fellowship. We acknowledge discussions with Vyoma Muralidhara in the early part of the project. We acknowledge Jens Chluba for comments on the draft and Rishi Khatri for discussions on some aspects of calculations.

\bibliography{ntsz}

\end{document}